\documentclass[preprint,prd,aps,showpacs,showkeys,nofootinbib]{revtex4}
\usepackage{graphicx}
\usepackage{dcolumn}
\usepackage{bm}
\usepackage{color}
\begin{document}
\title{The two-loop contributions to muon MDM in $U(1)_X$ SSM}

\author{Lu-Hao Su$^{1,2}$\footnote{suluhao0606@163.com}, Shu-Min Zhao$^{1,2}$\footnote{zhaosm@hbu.edu.cn}, Xing-Xing Dong$^{1,2}$\footnote{dxx$\_$0304@163.com}, Dan-Dan Cui$^{1,2}$, Tai-Fu Feng$^{1,2,3}$, Hai-Bin Zhang$^{1,2}$}

\affiliation{$^1$ Department of Physics, Hebei University, Baoding 071002, China}
\affiliation{$^2$ Key Laboratory of High-precision Computation and Application of Quantum Field Theory of Hebei Province, Baoding 071002, China}
\affiliation{$^3$ Department of Physics, Chongqing University, Chongqing 401331, China}
\begin{abstract}
  The MSSM is extended to the $U(1)_X$SSM, whose local gauge group is $SU(3)_C \times SU(2)_L \times U(1)_Y \times U(1)_X$. To obtain the $U(1)_X$SSM, we add  the new superfields to the MSSM, namely:
  three Higgs singlets $\hat{\eta},~\hat{\bar{\eta}},~\hat{S}$ and right-handed neutrinos $\hat{\nu}_i$. It can give light
  neutrino tiny mass at the tree level through the seesaw mechanism. The study of the contribution of the two-loop
  diagrams to the MDM of muon under $U(1)_X$SSM provides the possibility for us to search for new physics.
  In the analytical calculation of the loop diagrams (one-loop and two-loop diagrams),
  the effective Lagrangian method is used to derive muon MDM.
  Here, the considered two-loop diagrams include Barr-Zee type diagrams and rainbow type two-loop diagrams, especially
  $Z$-$Z$ rainbow two-loop diagram is taken into account.
  The obtained numerical results can reach  $7.4\times10^{-10}$, which can remedy the deviation between SM prediction and experimental data
  to some extent.
\end{abstract}


\maketitle
\section{Introduction}
In the development of quantum field theory, the study of lepton anomalous magnetic dipole moment(MDM) is very important.
The most accurate calculation of the lepton MDM is helpful in finding new physics beyond the standard model(SM) and Schwinger propose the electron
magnetic dipole moment (MDM) for the first time. It has been recognized that the lepton MDMs can provide
accurate testing of quantum electrodynamics (QED) \cite{1948Phys}. Therefore, it is of special significance to study the MDM of lepton.

At present, the experimental accuracy of measuring the MDM of the $\mu$, $a_\mu$ has reached 540ppb at the muon E821 anomalous magnetic moment measurement at Brookhaven National Laboratory (BNL) \cite{experiment}. And currently there are about 3.7 $\sigma$ \cite{muon1, muon2} between SM prediction and experimental measurement. In this precision, such measurements do more than testing the leading QED contributions \cite{test1}, meanwhile, the influence of the weak interaction and strong interaction of the SM of particle physics can be tested \cite{test2, test3, test4, test5}. The discovery of the Higgs made the SM a huge success \cite{muon1, higgs}.
Although the contribution from QED \cite{test1} plays an important role in lepton MDM, it is not the only factor. The hadronic contributions \cite{had1, had2} are also particularly important, and it is modified by hadron vacuum polarization and light-by-light \cite{lbl1, lbl2, lbl3} scattering contributions. Moreover, weak interaction \cite{w1, w2, w3} also has a certain effect on MDMs \cite{muon1, higgs}. As a result, the MDM of lepton can be expressed as \cite{muon1, mdm1, mdm2}:
\begin{eqnarray}
&&a_l^{SM}=a_l^{QED}+a_l^{EW}+a_l^{HAD}.
\end{eqnarray}

Due to the deficiency of MSSM which can not explain neutrino mass and solve $\mu$ problem, U(1) extension of MSSM is carried out. There are two U(1) groups in $U(1)_X$SSM: $U(1)_Y$ and $U(1)_X$, and we use SARAH software packages \cite{extend1, extend2, extend3} to study $U(1)_X$SSM.
On the basis of the MSSM, the superfields are added; then one obtains the additional Higgs, neutrino and gauge fields, but also corresponding
superpartners that extend the neutralino and sfermion sectors.
The CP-even parts of the three Higgs singlet fields $\eta$, $\overline{\eta}$, $S$ mix with the neutral CP-even parts of the two doublets $H_d$ and $H_u$ to form a tree order $5\times5$ CP-even Higgs mass matrix.
$m_{h0}$ is the tree level mass of the lightest CP-even Higgs in $U(1)_X$SSM, and it can be greater
than the corresponding mass at tree order in MSSM.
Therefore, the loop graph correction to $m_{h_0}$ in $U(1)_X$SSM needs not be very large.
In $U(1)_X$SSM, there are an $8\times8$ mass matrix for neutralinos and a $6\times6$ mass matrix for scalar neutrinos \cite{matrix}.
Next, we calculate the contribution of the two-loop diagrams to muon MDM under the $U(1)_X$SSM with the effective Lagrangian method.
The error between the experimental value and the predicted value by SM is \cite{mdm2, muon2, deviation3}
\begin{eqnarray}
&&\Delta a_\mu=a_\mu^{exp}-a_\mu^{SM}=(274\pm73)\times10^{-11}.
\end{eqnarray}
 In BLMSSM, the one-loop corrections are similar to the MSSM results \cite{m1, m2, m3, m4} in analytic form \cite{one}. The two-loop Barr-Zee type diagrams with a closed scalar (Fermi) loop between vector Boson and Higgs are studied in the frame work of BLMSSM \cite{one, two1, two2}. There are other works  \cite{other1, other2, other3} for the two-loop supersymmetric corrections \cite{s1, susy1, susy2} to the lepton MDM.

In this paper, the new physics contributions at one loop level are similar to the MSSM results in analytic form.
 The differences are that the mass matrices of scalar leptons, scalar neutrinos, neutrino, chargino and neutralino
 possess new parameters $g_X, v_{\bar{\eta}}, v_{\eta}$ and so on. The mass eigenstates of neutralinos, scalar neutrinos, neutrinos are more than those in the MSSM.
  Because the masses of those virtual fields
 ($W^{\pm}, Z, Z'$ gauge bosons, neutral and charged Higgs, as well as neutralinos and charginos) are much heavier than the muon mass $m_\mu$,
 we can expand the denominator corresponding to the ratio (external momentum to internal momentum) to simplify the loop calculation \cite{lepton}.
 We use the formula:
\begin{eqnarray}
\frac{1}{(k-p)^2-m_s^2}=\frac{1}{k^2-m_s^2}(1+\frac{2k\cdot p-p^2}{k^2-m_s^2}+\frac{4(k\cdot p)^2}{(k^2-m_s^2)^2})+  \dots~.
\end{eqnarray}
Here, p is the external momentum of $m_\mu$, and k is the internal momentum at the order of $m_{SUSY}$.
So, $\frac{m_\mu}{m_{SUSY}}$ $\ll$ 1 and $\frac{p}{k}$ $\ll$ 1.

  Our two-loop self energy diagrams contributing to the lepton MDMs are shown in Fig. 2. The two-loop triangle diagrams for $\mu\rightarrow \mu+\gamma$ can be obtained from the
  two-loop self energy diagrams by attaching a photon on the internal lines in all possible ways. The sum of all the two-loop triangle diagrams
  generated from a two-loop self energy diagram satisfies Ward-identity.

   The researched two-loop self energy diagrams include Barr-Zee type diagrams and rainbow type diagrams with Fermion sub-loop. Here, we suppose the
   scalar leptons and scalar quarks are very heavy, whose contributions from the studied two-loop diagrams can be neglected. In the works \cite{experiment, one, two1, two2, other1, other2, other3}, the two-loop rainbow diagrams
   with two vector bosons $Z$-$Z$ are not considered. However, in our work \cite{our} the order analysis of two-loop SUSY corrections \cite{susy1, susy2} to lepton MDM shows that
   the contributions from two-loop rainbow diagrams with $Z$-$Z$ are at the same order of the contributions from two-loop rainbow diagrams with $W$-$W$.
   Therefore, we take into account the $Z$-$Z$ rainbow diagrams.

  In the following, we introduce the specific form of $U(1)_X$SSM and its superfields.
 The analytic results of the one-loop corrections and two-loop corrections in the $U(1)_X$SSM are deduced in the section 3.
  Section 4 is used for the numerical calculation and discussion.
  In the last section, we have a special summary and discussion. Some mass matrics and Feynman rules are collected in the appendix.

\section{the $U(1)_X$ SSM}
The gauge group of the $U(1)_X$SSM is $SU(3)_C\otimes
SU(2)_L \otimes U(1)_Y\otimes U(1)_X$. To obtain the $U(1)_X$SSM, the MSSM is added with
three Higgs singlets $\hat{\eta},~\hat{\bar{\eta}},~\hat{S}$ and right-handed neutrinos $\hat{\nu}_i$. It can give light
neutrino mass at the tree level through the seesaw mechanism. The neutral CP-even parts of
$H_u,~ H_d,~\eta,~\bar{\eta}$ and $S$ mix together, forming $5\times5 $ mass squared matrix.
Because of the right handed neutrinos,
the mass matrix of neutrino is expended to $6\times6$. At the same time, the squared mass matrix of scalar
neutrinos turns to $6\times6$ too. For details of the mass matrix of particles, please see the appendix.

The superpotential for this model reads as:
\begin{eqnarray}
&&W=l_W\hat{S}+\mu\hat{H}_u\hat{H}_d+M_S\hat{S}\hat{S}-Y_d\hat{d}\hat{q}\hat{H}_d-Y_e\hat{e}\hat{l}\hat{H}_d+\lambda_H\hat{S}\hat{H}_u\hat{H}_d
\nonumber\\&&+\lambda_C\hat{S}\hat{\eta}\hat{\bar{\eta}}+\frac{\kappa}{3}\hat{S}\hat{S}\hat{S}+Y_u\hat{u}\hat{q}\hat{H}_u+Y_X\hat{\nu}\hat{\bar{\eta}}\hat{\nu}
+Y_\nu\hat{\nu}\hat{l}\hat{H}_u.
\end{eqnarray}

There are two Higgs doublets and three Higgs singlets. Their specific forms are shown below,
\begin{eqnarray}
&&H_{u}=\left(\begin{array}{c}H_{u}^+\\{1\over\sqrt{2}}\Big(v_{u}+H_{u}^0+iP_{u}^0\Big)\end{array}\right),
~~~~~~
H_{d}=\left(\begin{array}{c}{1\over\sqrt{2}}\Big(v_{d}+H_{d}^0+iP_{d}^0\Big)\\H_{d}^-\end{array}\right),
\nonumber\\
&&\eta={1\over\sqrt{2}}\Big(v_{\eta}+\phi_{\eta}^0+iP_{\eta}^0\Big),~~~~~~~~~~~~~~~
\bar{\eta}={1\over\sqrt{2}}\Big(v_{\bar{\eta}}+\phi_{\bar{\eta}}^0+iP_{\bar{\eta}}^0\Big),\nonumber\\&&
\hspace{4.0cm}S={1\over\sqrt{2}}\Big(v_{S}+\phi_{S}^0+iP_{S}^0\Big).
\end{eqnarray}
$v_u,~v_d,~v_\eta$,~ $v_{\bar\eta}$ and $v_S$ are the corresponding  VEVs of the Higgs superfields $H_u$, $H_d$, $\eta$, $\bar{\eta}$ and $S$.
Here, we define $\tan\beta=v_u/v_d$ and $\tan\beta_\eta=v_{\bar{\eta}}/v_{\eta}$. The definition of
$\tilde{\nu}_L$ and $\tilde{\nu}_R$ is
\begin{eqnarray}
\tilde{\nu}_L=\frac{1}{\sqrt{2}}\phi_l+\frac{i}{\sqrt{2}}\sigma_l,~~~~~~~~~~\tilde{\nu}_R=\frac{1}{\sqrt{2}}\phi_R+\frac{i}{\sqrt{2}}\sigma_R.
\end{eqnarray}

The soft SUSY breaking terms are
\begin{eqnarray}
&&\mathcal{L}_{soft}=\mathcal{L}_{soft}^{MSSM}-B_SS^2-L_SS-\frac{T_\kappa}{3}S^3-T_{\lambda_C}S\eta\bar{\eta}
+\epsilon_{ij}T_{\lambda_H}SH_d^iH_u^j\nonumber\\&&
-T_X^{IJ}\bar{\eta}\tilde{\nu}_R^{*I}\tilde{\nu}_R^{*J}
+\epsilon_{ij}T^{IJ}_{\nu}H_u^i\tilde{\nu}_R^{I*}\tilde{l}_j^J
-m_{\eta}^2|\eta|^2-m_{\bar{\eta}}^2|\bar{\eta}|^2\nonumber\\&&
-m_S^2S^2-(m_{\tilde{\nu}_R}^2)^{IJ}\tilde{\nu}_R^{I*}\tilde{\nu}_R^{J}
-\frac{1}{2}\Big(M_S\lambda^2_{\tilde{X}}+2M_{BB^\prime}\lambda_{\tilde{B}}\lambda_{\tilde{X}}\Big)+h.c~~.
\end{eqnarray}

With the singlet superfield $\hat{S}$ coupling to heavy fields in the most general way,
radiative corrections can induce very large terms in the effective action.
These terms are linear in $\hat{S}$ in the superpotential or linear in $S$ in $\mathcal{L}_{soft}$, and
they are called tadpole terms \cite{tt}. If they are too large, a tadpole problem appears in the model.
In the case of gauge mediated supersymmetry breaking(GMSB) \cite{gmsb}, the messenger fields $\hat{\varphi}$ with SM gauge quantum numbers and
supersymmetric mass terms($M_{mess}$) are the source of supersymmetry breaking.
The real and imaginary scalar components of the messenger fields have
different masses, and they are split by a scale $\hat{m}$.
This kind of supersymmetry breaking can be denoted as F-type splitting and represented by a F component of a
spurion superfield coupling to the messenger fields $\hat{\varphi}$.
Then, we obtain $l_W\sim C_S\hat{m}^2$ and $L_S\sim  C_S\hat{m}^4/M_{mess}$ with the simplest coupling $C_S \hat{S} \hat{\varphi}\hat{\varphi}$.
Here, $C_S$ is the small Yukawa coupling. There is not tadpole problem with $M_{mess}$ and the F-type splitting $\hat{m}$ not much larger than the weak scale \cite{ft}.
If these scales are larger than the weak scale, small Yukawa couplings can suppress the tadpole diagrams.

There is conflict between domain wall and tadpole problems, whose solutions are studied by Refs. \cite{z3}.
One can impose constraints on $Z_3$-symmetry breaking non-renormalisable interactions or
hidden sectors in the form of various additional symmetries \cite{z3s}.  As the tadpole terms, $Z_3$-symmetry
breaking renormalisable terms are generated radiatively and have very small coefficients. $Z_3$-symmetry breaking terms can  solve the domain
wall problem.

We use $Y^Y$ for  the $U(1)_Y$ charge and $Y^X$ for the $U(1)_X$ charge.
According to the textbook \cite{text}, the SM is
anomaly free.
The anomalies of $U(1)_X$SSM are more complicated than those of SM \cite{higgs}.
In the end, this model is anomaly free.
The presence of two Abelian groups $U(1)_Y$ and $U(1)_X$ in $U(1)_X$SSM have a new effect absent in the MSSM with just one Abelian gauge group $U(1)_Y$:
the gauge kinetic mixing. This effect can also be induced through RGEs, even if it is set to zero at $M_{GUT}$.
The covariant derivatives of this model have the general form \cite{model1, model2, model3}
\begin{eqnarray}
%%%%%%%%%%%%%%%%%%%%%%%%%%%%%%%%%%%%%%%%%%%%%%%%%%%%%%%%%%%%%%%%%%%%
&&D_\mu=\partial_\mu-i\left(\begin{array}{cc}Y,&X\end{array}\right)
\left(\begin{array}{cc}g_{Y},&g{'}_{{YX}}\\g{'}_{{XY}},&g{'}_{{X}}\end{array}\right)
\left(\begin{array}{c}A_{\mu}^{\prime Y} \\ A_{\mu}^{\prime X}\end{array}\right)\;.
%%%%%%%%%%%%%%%%%%%%%%%%%%%%%%%%%%%%%%%%%%%%%%%%%%%%%%%%%%%%%%%%%%%%
\label{gauge1}
\end{eqnarray}

Here, $A_{\mu}^{\prime Y}$ and $A^{\prime X}_\mu$ signify the gauge fields of $U(1)_Y$ and $U(1)_X$. We can do a basis conversion,
because the two Abelian gauge groups are unbroken.
The following formula can be obtained by using the appropriate matrix $R$ \cite{model1, model3}
\begin{eqnarray}
%%%%%%%%%%%%%%%%%%%%%%%%%%%%%%%%%%%%%%%%%%%%%%%%%%%%%%%%%%%%%%%%%%%%
&&\left(\begin{array}{cc}g_{Y},&g{'}_{{YX}}\\g{'}_{{XY}},&g{'}_{{X}}\end{array}\right)
R^T=\left(\begin{array}{cc}g_{1},&g_{{YX}}\\0,&g_{{X}}\end{array}\right)\;.
%%%%%%%%%%%%%%%%%%%%%%%%%%%%%%%%%%%%%%%%%%%%%%%%%%%%%%%%%%%%%%%%%%%%
\label{gauge2}
\end{eqnarray}

We deduce $\sin^2\theta_{W}^\prime$ as
\begin{eqnarray}
\sin^2\theta_{W}'=\frac{1}{2}-\frac{(g_{{YX}}^2-g_{1}^2-g_{2}^2)v^2+
4g_{X}^2\xi^2}{2\sqrt{(g_{{YX}}^2+g_{1}^2+g_{2}^2)^2v^4+8g_{X}^2(g_{{YX}}^2-g_{1}^2-g_{2}^2)v^2\xi^2+16g_{X}^4\xi^4}},
\end{eqnarray}
with $\xi=\sqrt{v_\eta^2+v_{\bar{\eta}}^2}$.
The new mixing angle $\theta_{W}^\prime$ appears in the couplings involving $Z$ and $Z^{\prime}$.

\section{formulation}
 We use the effective Lagrangian method, and the Feynman amplitude can be expressed by these dimension 6 operators \cite{lepton}.
 The higher order operators such as the dimension 8 operators are suppressed by additional factor $\frac{m_{\mu}^2}{m_{SUSY}^2}$ $\sim$ ($10^{-7}$, $10^{-8}$) comparing with the dimension 6 operators, which are neglected.
\begin{eqnarray}
&&\mathcal{O}_1^{\mp}=\frac{1}{(4\pi)^2}\bar{l}(i\mathcal{D}\!\!\!\slash)^3\omega_{\mp}l,
\nonumber\\
&&\mathcal{O}_2^{\mp}=\frac{eQ_f}{(4\pi)^2}\overline{(i\mathcal{D}_{\mu}l)}\gamma^{\mu}
F\cdot\sigma\omega_{\mp}l,
\nonumber\\
&&\mathcal{O}_3^{\mp}=\frac{eQ_f}{(4\pi)^2}\bar{l}F\cdot\sigma\gamma^{\mu}
\omega_{\mp}(i\mathcal{D}_{\mu}l),\nonumber\\
&&\mathcal{O}_4^{\mp}=\frac{eQ_f}{(4\pi)^2}\bar{l}(\partial^{\mu}F_{\mu\nu})\gamma^{\nu}
\omega_{\mp}l,\nonumber\\&&
\mathcal{O}_5^{\mp}=\frac{m_l}{(4\pi)^2}\bar{l}(i\mathcal{D}\!\!\!\slash)^2\omega_{\mp}l,
\nonumber\\&&\mathcal{O}_6^{\mp}=\frac{eQ_fm_l}{(4\pi)^2}\bar{l}F\cdot\sigma
\omega_{\mp}l.
\end{eqnarray}
with $\mathcal{D}_{\mu}=\partial_{\mu}+ieA_{\mu}$ and $\omega_{\mp}=\frac{1\mp\gamma_5}{2}$. $F_{{\mu\nu}}$ is the electromagnetic field strength, and
$m_{_l}$ is the lepton mass. Therefore, the Wilson coefficients of the operators $\mathcal{O}_{2,3,6}^{\mp}$ in the effective Lagrangian are of interest and their dimensions are -2.
The lepton MDM is the combination of the Wilson coefficients $C^{\mp}_{2,3,6}$ and can be obtained from the following effective Lagrangian
\begin{eqnarray}
&&{\cal L}_{_{MDM}}={e\over4m_{l}}\;a_{l}\;\bar{l}\sigma^{\mu\nu}
l\;F_{{\mu\nu}}\label{adm}.
\end{eqnarray}
\subsection{The one-loop corrections}
In $U(1)_X$SSM, the masses of the neutralinos, neutrinos, scalar neutrinos and scalar charged leptons are all
adopted comparing with those in MSSM.
The one loop new physics contributions to muon MDM comes from the diagrams in Fig. 1.

\begin{figure}[t]
\begin{center}
\begin{minipage}[c]{0.8\textwidth}
\includegraphics[width=5.0in]{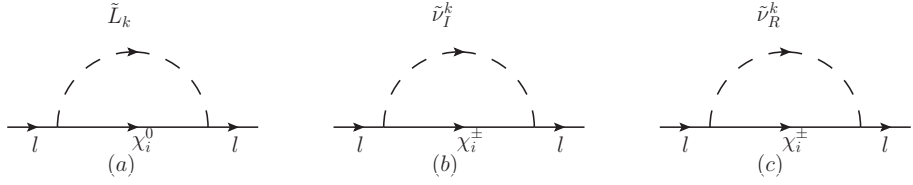}
\end{minipage}%
\caption{The one-loop self energy diagrams affect lepton MDMs in the $U(1)_X$SSM. The triangle diagrams
can be obtained by attaching a photon on the internal lines of the self energy diagrams in all possible ways.}
\end{center}
\end{figure}
In $U(1)_X$SSM, we find that our own results of the one-loop corrections are similar to the MSSM results in
analytic form. However, the mass matrices of scalar, Fermion and Majorana particles have relation with
new parameters $g_L, \bar{v}_L, v_L$ and so on.
The corrections to muon MDM from neutralinos and scalar leptons are expressed as
\begin{eqnarray}
&&a_{l}^{\tilde{L}\chi^{0}}=
-\sum_{i=1}^8\sum_{j=1}^6\Big[\Re(A_L^*A_R)
\sqrt{x_{\chi_j^{0}}x_{\mu}}x_{\tilde{L}_i}\frac{\partial^2 \mathcal{B}(x_{\chi_j^{0}},x_{\tilde{L}_i})}{\partial x_{\tilde{L}_i}^2}
\nonumber\\&&\hspace{1.4cm}+\frac{1}{3}(|A_L|^2+|A_R|^2)x_{\tilde{L}_i}x_{\mu}
\frac{\partial\mathcal{B}_1(x_{\chi_j^{0}},x_{\tilde{L}_i})}{\partial x_{\tilde{L}_i}}\Big].
\end{eqnarray}
where $x_M=\frac{M^2}{\Lambda^2}$, $M$ is the particle mass and $\Lambda$ is the mass scale.
The couplings $A_R,A_L$  are shown as
\begin{eqnarray}
&&A_R=\frac{1}{\sqrt{2}}g_1N_{i1}^{*}Z_{k2}^{E}+\frac{1}{\sqrt{2}}g_2N_{i2}^{*}Z_{k2}^{E}+\frac{1}{\sqrt{2}}g_{YX}N_{i5}^{*}Z_{k2}^{E}
-N_{i3}^{*}Y_\mu Z_{k5}^{E},\nonumber\\&&
A_L=-\frac{1}{\sqrt{2}}Z_{k5}^{E}(2g_1N_{i1}+(2g_{YX}+g_X)N_{i5})-Y_\mu^{*}Z_{k2}^EN_{i3}.
\end{eqnarray}

The matrices $Z^{E}$, $N$ respectively diagonalize the mass matrices of scalar lepton and neutralino.
The concrete forms of the functions $\mathcal{B}(x,y)$ and $\mathcal{B}_1(x,y)$ are
\begin{eqnarray}
\mathcal{B}(x,y)=\frac{1}{16 \pi
^2}\Big(\frac{x \ln x}{y-x}+\frac{y \ln
y}{x-y}\Big),~~~
\mathcal{B}_1(x,y)=(
\frac{\partial}{\partial y}+\frac{y}{2}\frac{\partial^2 }{\partial y^2})\mathcal{B}(x,y).
\end{eqnarray}
In a similar way, the corrections from chargino and CP-odd scalar neutrino are also obtained.
\begin{eqnarray}
&&a_{lI}^{\tilde{\nu}\chi^{\pm}}=\sum_{j=1}^2\sum_{k=1}^6
\Big[-2\Re(B_L^{*}B_R)\sqrt{x_{\chi_i^{-}}x_\mu}\mathcal{B}_1(x_{\tilde{\nu}_I}^{k},x_{\chi_i^{-}})
\nonumber\\&&\hspace{1.4cm}+\frac{1}{3}(|B_L|^2+|B_R|^2)x_\mu x_{\chi_i^{-}}\frac{\partial\mathcal{B}_1(x_{\tilde{\nu}_I}^{k},x_{\chi_i^{-}})}{\partial x_{\chi_i^{-}}}\Big].
\end{eqnarray}
Here, the $B_L$ and $B_R$ is
\begin{eqnarray}
B_L=-\frac{1}{\sqrt{2}}U_{j2}^{*}Z_{k2}^{I*}Y_\mu,~~~
B_R=\frac{1}{\sqrt{2}}g_2Z_{k2}^{I*}V_{j1}.
\end{eqnarray}
The corrections from chargino and CP-even scalar neutrino read as
\begin{eqnarray}
&&a_{lR}^{\tilde{\nu}\chi^{\pm}}=\sum_{j=1}^2\sum_{k=1}^6
\Big[-2\Re(C_L^{*}C_R)\sqrt{x_{\chi_i^{-}}x_\mu}\mathcal{B}_1(x_{\tilde{\nu}_R}^{k},x_{\chi_i^{-}})
\nonumber\\&&\hspace{1.4cm}+\frac{1}{3}(|C_L|^2+|C_R|^2)x_\mu x_{\chi_i^{-}}\frac{\partial\mathcal{B}_1(x_{\tilde{\nu}_R}^{k},x_{\chi_i^{-}})}{\partial x_{\chi_i^{-}}}\Big].
\end{eqnarray}
Here, the $C_L$ and $C_R$ are
\begin{eqnarray}
C_L=\frac{1}{\sqrt{2}}U_{j2}^{*}Z_{k2}^{R*}Y_\mu,~~~
C_R=-\frac{1}{\sqrt{2}}g_2Z_{k2}^{R*}V_{j1}.
\end{eqnarray}

\begin{figure}
  \setlength{\unitlength}{1mm}
  \centering
  \includegraphics[width=3.0in]{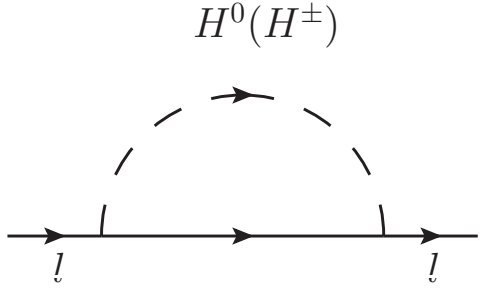}
  \caption[]{One-loop Higgs diagrams contributings to muon MDM in the $U(1)_X$SSM. It is suppressed by the square of the lepton Yukawa coupling.}\label{OLH}
\end{figure}

Here, $U,V$ are used to diagonalize the chargino mass matrix,
and the mass squared matrix of CP-even (CP-odd) scalar neutrino is diagonalized by $Z^{R}~(Z^I)$.
The one-loop Higgs contribution to muon MDM is shown by the Fig. \ref{OLH}, which
is suppressed by the factor $\frac{m_{l^I}^2}{m^2_W}\sim10^{-6}$. In numerical estimation, the correction from Higgs one-loop
diagram is around $10^{-13}$, and we can neglect it safely.
So, the one-loop corrections to lepton MDM can be expressed as
\begin{eqnarray}
\Delta a_l^{one-loop}=a_{l}^{\tilde{L}\chi^{0}}+a_{lI}^{\tilde{\nu}\chi^{\pm}}+a_{lR}^{\tilde{\nu}\chi^{\pm}}.
\end{eqnarray}

\subsection{The two-loop corrections}

 As discussed in Ref. \cite{our}, the Barr-Zee two-loop diagrams (Fig. 3 (a), (b), (c)) and rainbow two-loop diagrams (Fig. 3 (d), (g))
 have not small factors to muon MDM. That is to say, they have considerable contributions to muon MDM.
   The triangle diagrams can be obtained by attaching a photon on the internal lines of the self energy diagrams in all possible ways.
  The sum of all the triangle diagrams corresponding to one self energy diagram satisfy the Ward-identity and the CTP invariance.

\begin{figure}[t]
\begin{center}
\begin{minipage}[c]{1.0\textwidth}
\includegraphics[width=6.0in]{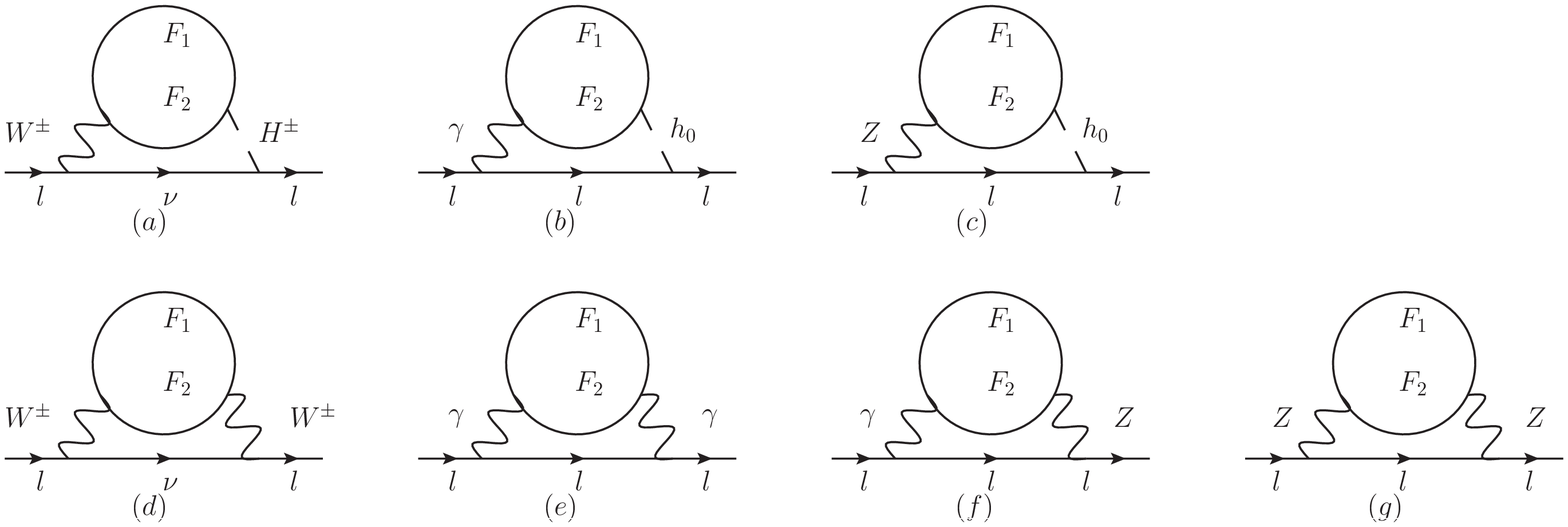}
\end{minipage}%
\caption{The two-loop Barr-Zee and rainbow type diagrams affect lepton MDMs in the $U(1)_X$SSM.}
\end{center}
\end{figure}

At first, we consider the corrections from Fig. 3(a). Under the assumption $m_F=m_{F_1}=m_{F_2}\gg m_W$, the results \cite{ffa} can be simplified as

\begin{eqnarray}
&&a_l^{WH}=\frac{G_F m_l m_W^2s_W}{128e\pi^4}\sum_{F_1=\chi^{\pm}}\sum_{F_2=\chi^0}\frac{H_{\bar{l}H\nu}^L}{ m_F}\Big\{\Big[\frac{21}{4}-\frac{5}{18}Q_{F_1}+(3+\frac{Q_{F_1}}{3})
(\ln{m_{F_1}^2}\nonumber\\
&&\qquad\quad\hspace{-0.8cm}-\varrho_{1,1}(m_W^2,m_{H^\pm}^2))\Big]\Re(H_{HF_1F_2}^LH_{WF_1F_2}^L+H_{HF_1F_2}^RH_{WF_1F_2}^R)
+\Big[\frac{19-20Q_{F_1}}{9}\nonumber\\
&&\qquad\quad\hspace{-0.8cm}+\frac{2-4Q_{F_1}}{3}
(\ln{m_{F_1}^2}-\varrho_{1,1}(m_W^2,m_{H^\pm}^2))\Big]\Re(H_{HF_1F_2}^LH_{WF_1F_2}^R+H_{HF_1F_2}^RH_{WF_1F_2}^L)
\nonumber\\
&&\qquad\quad\hspace{-0.8cm}+\Big[\hspace{-0.1cm}-\hspace{-0.1cm}\frac{16}{9}\hspace{-0.1cm}-\hspace{-0.1cm}\frac{2\hspace{-0.1cm}+\hspace{-0.1cm}6Q_{F_1}}{3}
(\ln{m_{F_1}^2}\hspace{-0.1cm}-\hspace{-0.1cm}\varrho_{1,1}(m_W^2,m_{H^\pm}^2))\Big]\Re(H_{HF_1F_2}^LH_{WF_1F_2}^L\hspace{-0.1cm}-\hspace{-0.1cm}H_{HF_1F_2}^RH_{WF_1F_2}^R)
\nonumber\\
&&\qquad\quad\hspace{-0.8cm}+\Big[\hspace{-0.1cm}-\hspace{-0.1cm}\frac{2Q_{F_1}}{9}\hspace{-0.1cm}-\hspace{-0.1cm}\frac{6\hspace{-0.1cm}-\hspace{-0.1cm}2Q_{F_1}}{3}
(\ln{m_{F_1}^2}\hspace{-0.1cm}-\hspace{-0.1cm}\varrho_{1,1}(m_W^2,m_{H^\pm}^2))\Big]\Re(H_{HF_1F_2}^LH_{WF_1F_2}^R\hspace{-0.1cm}-\hspace{-0.1cm}H_{HF_1F_2}^RH_{WF_1F_2}^L)\Big\}.
\end{eqnarray}
where $\varrho_{1,1}(x,y)=\frac{x\ln x-y\ln y}{x-y}$. $H_{HF_1F_2}^{L,R}$ and $H_{WF_1F_2}^{L,R}$ represent the coupling coefficients of the corresponding vertices with the presentation
\begin{eqnarray}
&&\mathcal{L}_{HF_1F_2}=i\bar{F_1}(H_{HF_1F_2}^{L}P_L+H_{HF_1F_2}^{R}P_R)F_2H^{\pm},\nonumber\\&&
 \mathcal{L}_{WF_1F_2}=i\bar{F_1}(H_{WF_1F_2}^{L}\gamma_\mu P_L+H_{WF_1F_2}^{R}\gamma_\mu P_R)F_2W^\mu.
\end{eqnarray}
Their concrete forms are collected in the appendix.

Then under the assumption $m_F=m_{F_1}=m_{F_2}\gg m_{h_0}$, the two-loop Barr-Zee type diagrams contributing to the lepton MDMs  corresponding to Figs. 3(b) and 3(c) can be simplified as
\begin{eqnarray}
&&a_l^{\gamma h_0}=\frac{G_FQ_fQ_{F_1}m_lm_W^2s_W^2}{16\pi^4}\sum_{F_1=F_2=\chi^\pm}\frac{1}{m_{F_1}}
\Re(H_{h_0F_1F_2}^L)\Big[1+\ln\frac{m_{F_1}^2}{m_{h_0}^2}\Big].
\nonumber\\&&a_l^{Zh_0}=\frac{\sqrt{2}m_l }{512\pi^4}\sum_{F_1=F_2=\chi^{\pm},\chi^0}
\frac{H_{h_0l\bar{l}}}{m_{F_1}}\Big[\varrho_{1,1}(m_Z^2,m_{h_0}^2)-\ln{m_{F_1}^2}-1\Big]
\nonumber\\&&\qquad\quad\times(H^L_{Zll}-H^R_{Zll})\Re(H_{h_0F_1F_2}^LH_{ZF_1F_2}^L+H_{h_0F_1F_2}^RH_{ZF_1F_2}^R).
\end{eqnarray}
$Q_f$ is the electric charge of the external lepton $m_\mu$. $Q_{F_1}$ and $Q_{F_2}$ are the electric charges of the internal charginos.

 Under the assumption $m_F=m_{F_1}=m_{F_2}\gg m_W\sim m_Z$, the two-loop rainbow type diagrams contributing to the lepton MDMs corresponding to Fig. 3(d),
3(e) and 3(f) can be simplified as
\begin{eqnarray}
&&a_l^{WW}=\frac{G_F m_l^2}{192\sqrt{2}\pi^4}\sum_{F_1=\chi^{\pm}}\sum_{F_2=\chi^0}\Big\{(18Q_{F_1}-13)(|H_{WF_1F_2}^L|^2+|H_{WF_1F_2}^R|^2)
\nonumber\\
&&\qquad\quad+3(Q_{F_1}-3)(|H_{WF_1F_2}^L|^2-|H_{WF_1F_2}^R|^2)+11\Re(H_{WF_1F_2}^{R*}H_{WF_1F_2}^L)\Big\}.
\nonumber\\&&a_l^{\gamma\gamma}=\frac{\sqrt{2}e^2G_FQ_{F_1}^2m_l^2}{180\pi^4}\sum_{F_1=F_2=\chi^\pm}\frac{m_W^2}{m_{F_1}^2}.
\nonumber\\&&a_l^{\gamma Z}=\frac{Q_{F_1}m_l^2e^2}{256\pi^4}(H^R_{Zll}-H^L_{Zll})\sum_{F_1=F_2=\chi^\pm}\frac{1}{m_{F_1}^2}
\Re(H_{ZF_1F_2}^L-H_{ZF_1F_2}^R)\Big[35+\ln\frac{m_{F_1}^2}{m_Z^2}\Big].
\end{eqnarray}

Here, we focus on $Z$-$Z$ two-loop rainbow diagram in the Fig. 3(g). In the references \cite{two1, two2, other1, other2, other3},
the $Z$-$Z$ two-loop diagram has not been taken into account. But in this paper,
we research the $Z$-$Z$ two-loop diagram and show the simplified result.
The triangle diagrams of the two-loop $Z$-$Z$ rainbow self-energy diagram can be divided into two parts:
 1 the external photon is attached on the virtual lepton between $Z$-$Z$, and the corresponding counter term
should be considered. After the simplification, the contributions from this type two-loop diagrams are tiny, and
can be neglected safely.
2 the external photon is attached on the charged fermion sub-loop. The full two-loop results of this type triangle diagrams
are very tedious and can be found in Ref. \cite{ffa}.
With the assumption $m_F = m_{F1} = m_{F2} \gg m_W \sim m_Z$, we simplify the tedious two-loop results to the order $\frac{m_\mu^2}{M_Z^2}$ $\sim$ $10^{-6}$ or $\frac{m_\mu^2}{m_{SUSY}^2}$, and obtain the concise form.
\begin{eqnarray}
&&a_{l}^{ZZ}=
-\frac{Q_{F_1} x_l}{1024\pi^4}\sum_{F_1=F_2=\chi^\pm}\Big\{
\Big(|H^L_{ZF_1F_2}|^2+|H^R_{ZF_1F_2}|^2\Big)\Big(|H^L_{Zll}|^2
+|H^R_{Zll}|^2\Big)[\frac{-12 \log x_F-30}{x_Z}]\nonumber\\&&
+\Big(|H^L_{ZF_1F_2}|^2-|H^R_{ZF_1F_2}|^2\Big)\Big(|H^L_{Zll}|^2
-|H^R_{Zll}|^2\Big)[\frac{-3 \log x_Z+3 \log x_F+2}{9 x_F}]
\nonumber\\&&+\Re(H^L_{ZF_1F_2}H^R_{ZF_1F_2})\Big(|H^L_{Zll}|^2
+|H^R_{Zll}|^2\Big)[\frac{-6 \log x_Z+6 \log x_F+4}{9 x_F}]\nonumber\\&&
+\Big(|H^L_{ZF_1F_2}|^2+|H^R_{ZF_1F_2}|^2\Big)H^L_{Zll}H^R_{Zll}
[16\frac{(\log x_F-\log x_Z) (\log x_F+2)+2}{x_Z}]\Big\}.
\end{eqnarray}
At two-loop level, the contributions to lepton MDMs can be summarized as
\begin{eqnarray}
&&\Delta a_l^{two-loop}=\Delta a_l^{one-loop}+a_l^{WH}+a_l^{\gamma h_0}+a_l^{Z h_0}+a_l^{WW}+a_l^{\gamma\gamma}+a_l^{\gamma Z}+a_{l}^{ZZ}.
\end{eqnarray}

\section{the numerical results}
In this section, we will discuss the numerical results. The lightest CP-even higgs mass is considered as an input parameter, which is $m_{h0}$=125.1 GeV \cite{hmass1, hmass2}.
The parameters used in $U(1)_X$SSM are given below:
\begin{eqnarray}
&&g_X=0.3,~g_{YX}=0.2,~\lambda_C=-0.3,~\kappa=0.3,~T_{\lambda_H}=1.8~{\rm TeV},~T_{\kappa}=1.6~{\rm TeV},~B_\mu=1~{\rm TeV^2},
\nonumber\\&&T_{\lambda_C}=M_{BL}=M_1=1~{\rm TeV},~v_{\eta}=15.5\times\cos{\beta_\eta}~{\rm TeV},
~v_{\bar{\eta}}=15.5\times\sin{\beta_\eta}~{\rm TeV},~\tan{\beta_\eta}=1,
\nonumber\\&&m_{SF}=2~{\rm TeV^2},~\mu_x=500~{\rm GeV},~M_{{BB}^\prime}=0.4~{\rm TeV},~tan{\beta}=11,~T_{E11}=T_{E22}=T_{E33}=1~{\rm TeV},
\nonumber\\&&T_{\nu33}=1600~{\rm GeV},~M_{\nu11}=M_{\nu22}=0.5~{\rm TeV^2},~M_{\nu33}=250^{2}~{\rm GeV^2}, ~ Y_{X11}=Y_{X22}=0.5, \nonumber\\&&M_{L11}=M_{E11}=M_{L22}=M_{E22}=M_{L33}=M_{E33}=2.4~{\rm TeV^2}, ~T_{X11}=T_{X22}=-1~{\rm TeV},
\nonumber\\&&T_{X33}=-2~{\rm TeV},~Y_{X33}=0.4,~ B_S=1~{\rm TeV^2},~ \lambda_H=0.1, ~l_W=5~{\rm TeV^2}.
\end{eqnarray}

To simplify the discussion, the parameters $Y_X$, $T_X$, $T_\nu$, $T_E$, $M_{L}$, $M_{E}$ and $M_{\nu}$ are supposed as diagonal matrices.
In the follow,  the remaining tunable parameters are
$M_2$, $M_S$, $\mu$, $v_S$, $M_{E11}=M_{E22}=M_{E33}=M_E$.
Next, we will analyze the effects of these parameters on the contributions to the muon MDM.

\begin{figure}[t]
\begin{center}
\begin{minipage}[c]{0.48\textwidth}
\includegraphics[width=2.9in]{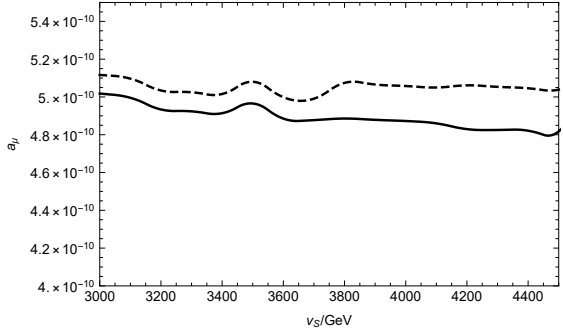}
\end{minipage}%
\caption{$a_\mu$ versus $v_S$. The solid (dashed) line corresponds to the result with $M_2=1000~(1200)~\rm{GeV}$.} \label{VS}
\end{center}
\end{figure}

With the parameters $M_S=1200~\rm{GeV}$, $\mu=300~\rm{GeV}$, $M_{E}=1~\rm{TeV^2}$, we plot $a_\mu$ versus $v_S$ in the Fig. \ref{VS}. $v_S$ is VEV of the Higgs singlet S, which
 affects the particle masses. So, it is expected that $v_S$ gives influence on $a_\mu$. The dashed curve corresponds to $M_2=1200~\rm{GeV}$, and the solid line corresponds to $M_2=1000~\rm{GeV}$. The both curves vary gently with $v_S$ in the region $(3000\sim4500)$ GeV. The dashed curve is a little larger than the solid curve.
 In the whole, the both curves are around the order $5.0\times 10^{-10}$. These corrections are considerable.

\begin{figure}[t]
\begin{center}
\begin{minipage}[c]{0.48\textwidth}
\includegraphics[width=2.9in]{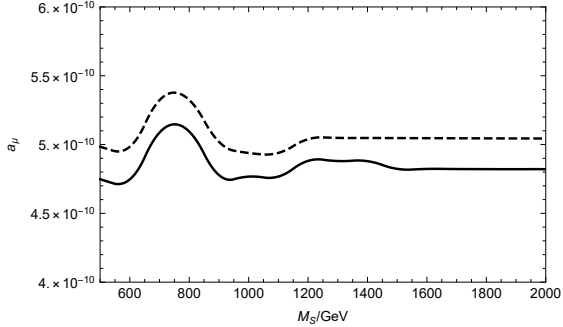}
\end{minipage}%
\caption{ $a_\mu$ versus $M_S$. The dashed (solid) line denotes the result with $M_2=1200~(800)~\rm{GeV}$.} \label{MS}
\end{center}
\end{figure}

$M_S$ is the mass of the new gaugino, and it appears in the mass matrix of neutralino.
Therefore, the one loop and two loop corrections relating with neutralino
are affected by $M_S$. Supposing $\mu=300~\rm{GeV}$, $v_S=4500~\rm{GeV}$, $M_{E}=1~\rm{TeV^2}$,
we show $a_\mu$ varying with $M_S$ by the solid curve ($M_2=800$ GeV) and dashed curve ($M_2=1200$ GeV) in the Fig.\ref{MS}.
During the region ($1000\sim1500$) GeV, the solid curve is almost an increasing function. As $M_S$ is near 750 GeV,
the biggest value of the dashed line can reach $5.4\times10^{-10}$.
As $M_S$ is larger than 1200 GeV, the dashed line is almost horizontal and about $5.0\times10^{-10}$.
The smallest value of the dashed line is about $4.8\times10^{-10}$. Generally speaking, $M_S$ is a sensitive parameter and influence $a_\mu$
obviously. The biggest value of the dashed line is about $0.3\times10^{-10}$ larger than the horizon. The change in  $M_S$ has an effect on the two mass eigenvalues of neutralino $\chi^0$.
The mass matrix of neutralino affects the contributions from the $W$-$H$ two-loop diagram, the $Z$-$h_0$ two-loop diagram, the $W$-$W$ two-loop diagram,
the $Z$-$Z$ two-loop diagram, like Fig. 3 (a), (c), (d), (g). Among them, the influence on $W$-$W$ two-loop Fig. 3 (d) is greater.
The contributions of these two-loop diagrams add up to the large bump.
The condition of the solid line is similar as that of the the dashed line.

\begin{figure}[t]
\begin{center}
\begin{minipage}[c]{0.48\textwidth}
\includegraphics[width=2.9in]{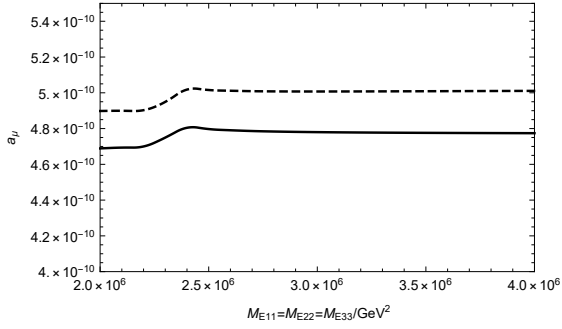}
\end{minipage}%
\caption{The relationship between $a_\mu$ and $M_{E11}=M_{E22}=M_{E33}=M_E$. The dashed line expresses $M_2=1000~\rm{GeV}$, and the solid line expresses $M_2=800~\rm{GeV}$.} \label{ME}
\end{center}
\end{figure}

Then, we analyze the effects of the parameters $M_{E}$ on the results and try to find their reasonable ranges.
The parameters $M_{E}$ are the diagonal elements of the mass squared matrix of the charged slepton, and they are major factors for slepton mass.
Based on $M_S=1200~\rm{GeV}$, $\mu=300~\rm{GeV}$, $v_S=3000~\rm{GeV}$, the numerical results are shown by the dashed curve and solid curve corresponding to $M_2=1000$ GeV and
$M_2=800$ GeV respectively. In the Fig.\ref{ME}, $a_\mu$ varies with $M_E$ in the range from $2.0~\rm{TeV^2}$ to $4.0~\rm{TeV^2}$.
The both curves possess similar behavior, and the dashed curve is above the solid curve. In the $M_E$ region ($2.0~\rm{TeV^2}$ to $2.5~\rm{TeV^2}$),
the both curves are slowly increasing functions.  As $2.5~\rm{TeV^2}<M_E<4.0~\rm{TeV^2}$, they hardly changes at all.

\begin{figure}[t]
\begin{center}
\begin{minipage}[c]{0.48\textwidth}
\includegraphics[width=2.9in]{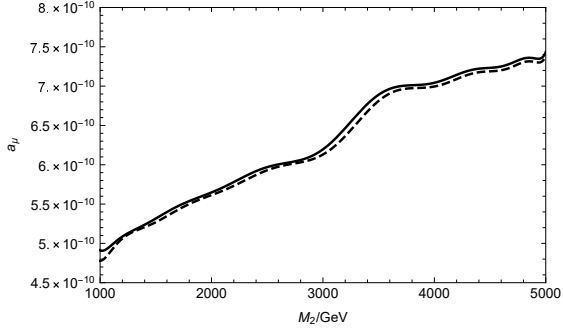}
\end{minipage}%
\caption{The effect of $M_2$ on the results. The dashed line expresses $v_S=3800\rm{GeV}$, and the solid line expresses $v_S=4300\rm{GeV}$.} \label{M2}
\end{center}
\end{figure}

The SU(2) gaugino mass $M_2$ is the diagonal element of chargino and neutralino, so $M_2$
gives influence to the masses of chargino and neutralino. Certainly,
the corrections to $a_\mu$ are affected by $M_2$. From the Figs. (\ref{VS}, \ref{MS}, \ref{ME}), one can find that $M_2$
is a sensitive parameter. Adopting $M_S=1200~\rm{GeV}$, $\mu=300~\rm{GeV}$, $M_{E}=1~\rm{TeV^2}$, we plot the results by the dashed curve ($v_S=3800~\rm{GeV}$) and solid curve
($v_S=4300~\rm{GeV}$) in the Fig.\ref{M2}.
The both curves are similar and almost overlap, which increase obviously with the enlarging $M_2$.
When $M_2$ is near 5000 GeV, the both curves can reach $7.4\times10^{-10}$ about one $\sigma$ deviation between experiment data and SM prediction.

\begin{figure}[t]
\begin{center}
\begin{minipage}[c]{0.48\textwidth}
\includegraphics[width=2.9in]{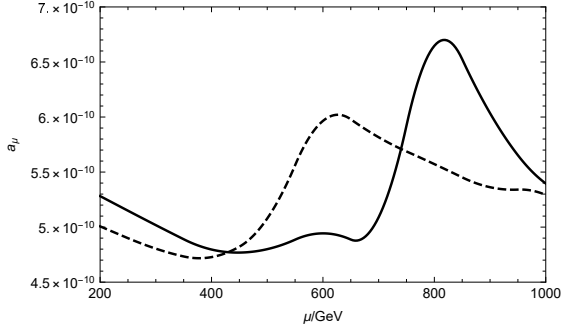}
\end{minipage}%
\caption{The influence of $\mu$ on $a_\mu$. The dashed line expresses $M_2=800~\rm{GeV}$, and the solid line expresses $M_2=1000~\rm{GeV}$.} \label{mu}
\end{center}
\end{figure}

To see how $\mu$ affects the muon MDM, we suppose $M_S=1200~\rm{GeV}$, $v_S=3000~\rm{GeV}$, $M_{E}=1~\rm{TeV^2}$ and plot the results versus $\mu$. The solid (dashed) curve
corresponds to the results obtained with $M_2=1000(800)$ GeV, which is shown in the Fig.\ref{mu}.
It is obvious that, $\mu$ affects $a_\mu$ strongly. Their smallest value is about $4.7\times10^{-10}$ and largest value is about $6.8\times10^{-10}$.
When $\mu$ ia larger than 800 GeV, the both curves decrease.
The dashed line expresses the result with $M_2$ = 800 GeV, and it comes up the large bump ($a_\mu$ = 6.01 $\times$ $10^{-10}$) around $\mu$ = 630 GeV.
The change in $\mu$ affects the mass matrices of chargino and neutralino. Their mass eigenvalues and turning matrices can influence the contributions of
all the studied two-loop diagrams in the Fig. 3. So the effect from $\mu$ is  larger than that of $M_S$.
$\mu$ is a sensitive parameter and affects the result obviously in the Fig. 8.

\section{discussion and conclusion}

 $U(1)_X$SSM is the $U(1)_X$ extension of MSSM, and it has new superfields.
 Using the effective Lagrangian method, we calculate and analyze the contributions from
 the one-loop and two-loop diagrams to muon MDM in $U(1)_X$ SSM.
The constraint from the lightest CP-even Higgs with the top-scalar-top loop correction is taken into account.
As introduced in the section II, these new particles lead to new corrections to $a_\mu$.
 The studied contributions are composed of one-loop diagrams, two-loop Barr-Zee type diagrams and two-loop rainbow diagrams
\begin{eqnarray}
&&\Delta a_l^{two-loop}=\Delta a_l^{one-loop}+a_l^{WH}+a_l^{\gamma h_0}+a_l^{Z h_0}+a_l^{WW}+a_l^{\gamma\gamma}+a_l^{\gamma Z}+a_{l}^{ZZ}.
\label{twoloop}
\end{eqnarray}

The two-loop rainbow diagrams with two $Z$ bosons are not considered in the works \cite{experiment, two1, two2, other1, other2, other3}. In our numerical calculation,
 the contributions from two-loop rainbow diagrams with two $Z$ bosons are at the same order of the contributions from the
 two-loop rainbow diagrams with two $W$ bosons. This result is consistent with the two-loop order analysis in our previous work \cite{our}.
 Therefore, the contributions from the two-loop rainbow diagrams with two $Z$ bosons are not small, and they should be taken into account.

The numerical results show that the parameters $\mu$, $M_2$, $M_S$, $v_S$, $M_{E11}=M_{E22}=M_{E33}$ are all influential to $a_\mu$.
$M_2$ is a sensitive parameter, and the largest value of $a_\mu$ in our used parameter space can reach $7.4\times 10^{-10}$.
In the large part of the parameter space,
the contributions from these loop diagrams are about $5.0\times 10^{-10}$. These corrections can make up the deviation between
experimental data and SM prediction to about one quarter. As we all know, there are a lot of two-loop diagrams contributing to muon MDM.
Some non-studied two-loop diagrams can also give important corrections to muon MDM, which can improve theoretical value more.
Because the calculation of the two-loop diagrams are very tedious, we will study  other two-loop diagrams in the future work.

\section{acknowledgments}
This work is supported by National Natural Science Foundation of China(NNSFC)(Nos. 11535002, 11705045), Natural Science Foundation of Hebei Province (A2020201002) and the youth top-notch talent support program of the Hebei Province.

\appendix
\begin{center}
\Large{{\bf Appendix }}
\end{center}
\vspace{-8mm}
~\\
The mass matrix for slepton with the basis $(\tilde{e}_L,\tilde{e}_R)$
\begin{eqnarray}
m_{\tilde{e}}^2=
\left({\begin{array}{*{20}{c}}
m_{\tilde{e}_{L}\tilde{e}_{L}^*} & \frac{1}{2}(\sqrt{2}v_dT_e^\dag - v_u(\lambda_H\chi S + \sqrt{2}\mu)Y_e^\dag) \\
\frac{1}{2}(\sqrt{2}v_dT_e - v_uY_e(\sqrt{2}\mu^* + \chi S\lambda_H^*)) & m_{\tilde{e}_{R}\tilde{e}_{R}^*} \\
\end{array}}
\right),
\end{eqnarray}
\begin{eqnarray}
&& m_{\tilde{e}_{L}\tilde{e}_{L}^*}=m_{\tilde{l}}^{2}+\frac{1}{8}\Big((g_{1}^{2}+g_{YX}^{2}+g_{YX}g_{X}-g_{2}^{2})(v_d^2-v_u^2)
+2g_{YX}g_{X}(v_\eta^2-v_{\bar{\eta}}^2)\Big)+\frac{1}{2}v_d^2Y_e^{\dag} Y_e,
\nonumber\\&& m_{\tilde{e}_{R}\tilde{e}_{R}^*}=m_e^2-\frac{1}{8}\Big([2(g_1^2+g_{YX})+3g_{YX}g_X+g_X^2](v_d^2-v_u^2)
+(4g_{YX}g_X+2g_X^2)(v_\eta^2-v_{\bar{\eta}}^2)\Big)
\nonumber\\&&\hspace{1.8cm}+\frac{1}{2}v_d^2Y_eY_e^{\dag}.
\end{eqnarray}
The mass matrix for CP-even sneutrino $({\phi}_{l}, {\phi}_{r})$ reads
\begin{eqnarray}
m^2_{\tilde{\nu}^R} = \left(
\begin{array}{cc}
m_{{\phi}_{l}{\phi}_{l}} &m^T_{{\phi}_{r}{\phi}_{l}}\\
m_{{\phi}_{l}{\phi}_{r}} &m_{{\phi}_{r}{\phi}_{r}}\end{array}
\right),
\end{eqnarray}
\begin{eqnarray}
&&m_{{\phi}_{l}{\phi}_{l}}= \frac{1}{8} \Big((g_{1}^{2} + g_{Y X}^{2} + g_{2}^{2}+ g_{Y X} g_{X})( v_{d}^{2}- v_{u}^{2})
+  g_{Y X} g_{X}(2 v_{\eta}^{2}-2 v_{\bar{\eta}}^{2})\Big)
\nonumber\\&&\hspace{1.8cm}+\frac{1}{2} v_{u}^{2}{Y_{\nu}^{T}  Y_\nu}  + m_{\tilde{L}}^2,
 %%%%%%%%%%%%%%%%%%%%%%%%%%%%%%
 \\&&m_{{\phi}_{l}{\phi}_{r}} = \frac{1}{\sqrt{2} } v_uT_\nu  +  v_u v_{\bar{\eta}} {Y_X  Y_\nu}
  - \frac{1}{2}v_d ({\lambda}_{H}\chi S  + \sqrt{2} \mu )Y_\nu,\\&&
m_{{\phi}_{r}{\phi}_{r}}= \frac{1}{8} \Big((g_{Y X} g_{X}+g_{X}^{2})(v_{d}^{2}- v_{u}^{2})
+2g_{X}^{2}(v_{\eta}^{2}- v_{\bar{\eta}}^{2})\Big) + v_{\eta} \chi S Y_X {\lambda}_{C}\nonumber \\&&\hspace{1.8cm}
 +m_{\tilde{\nu}}^2 + \frac{1}{2} v_{u}^{2}|Y_\nu|^2+  v_{\bar{\eta}} (2 v_{\bar{\eta}}|Y_X|^2  + \sqrt{2} T_X).
\end{eqnarray}
The mass matrix for CP-odd sneutrino $({\sigma}_{l}, {\sigma}_{r})$ is also deduced here
\begin{eqnarray}
m^2_{\tilde{\nu}^I} = \left(
\begin{array}{cc}
m_{{\sigma}_{l}{\sigma}_{l}} &m^T_{{\sigma}_{r}{\sigma}_{l}}\\
m_{{\sigma}_{l}{\sigma}_{r}} &m_{{\sigma}_{r}{\sigma}_{r}}\end{array}
\right),
\end{eqnarray}
\begin{eqnarray}
&&m_{{\sigma}_{l}{\sigma}_{l}}= \frac{1}{8} \Big((g_{1}^{2} + g_{Y X}^{2} + g_{2}^{2}+  g_{Y X} g_{X})( v_{d}^{2}- v_{u}^{2})
+  2g_{Y X} g_{X}(v_{\eta}^{2}-v_{\bar{\eta}}^{2})\Big)
\nonumber\\&&\hspace{1.8cm}+\frac{1}{2} v_{u}^{2}{Y_{\nu}^{T}  Y_\nu}  + m_{\tilde{L}}^2,
 %%%%%%%%%%%%%%%%%%%%%%%%%%%%%%
 \\&&m_{{\sigma}_{l}{\sigma}_{r}} = \frac{1}{\sqrt{2} } v_uT_\nu -  v_u v_{\bar{\eta}} {Y_X  Y_\nu}
  - \frac{1}{2}v_d ({\lambda}_{H}\chi S  + \sqrt{2} \mu )Y_\nu,\\&&
m_{{\sigma}_{r}{\sigma}_{r}}= \frac{1}{8} \Big((g_{Y X} g_{X}+g_{X}^{2})(v_{d}^{2}- v_{u}^{2})
+2g_{X}^{2}(v_{\eta}^{2}- v_{\bar{\eta}}^{2})\Big)- v_{\eta} \chi S Y_X {\lambda}_{C}\nonumber \\&&\hspace{1.8cm}
+m_{\tilde{\nu}}^2 + \frac{1}{2} v_{u}^{2}|Y_\nu|^2+  v_{\bar{\eta}} (2 v_{\bar{\eta}}Y_X  Y_X  - \sqrt{2} T_X).
\end{eqnarray}
Mass matrix for charginos in the basis:($\tilde{W}^-$,$\tilde{H}_d^-$),($\tilde{W}^+$,$\tilde{H}_u^+$)
\begin{eqnarray}
m_{{\tilde{\chi}}^-}=
\left({\begin{array}{*{20}{c}}
M_2 & \frac{1}{\sqrt{2}}g_2v_u \\
\frac{1}{\sqrt{2}}g_2v_d & \frac{1}{\sqrt{2}}\lambda_H\chi S+\mu \\
\end{array}}
\right),
\end{eqnarray}
The matrix is diagonalized by U and V
\begin{eqnarray}
U^*m_{{\tilde{\chi}}^-}V^\dag = m_{{\tilde{\chi}}^-}^{dia}.
\end{eqnarray}
The mass matrix for charged Higgs in the basis:($H_d^{-}$,$H_u^{+,*}$),($H_d^{-,*}$,$H_u^{+}$)
\begin{eqnarray}
m_{H_-}^2=
\left({\begin{array}{*{20}{c}}
m_{{H_d^{-}}H_d^{-,*}} & m_{H_u^{+,*}H_d^{-,*}}^{*} \\
m_{H_d^{-}H_u^{+}} & m_{H_u^{+,*}H_u^{+}} \\
\end{array}}
\right),
\end{eqnarray}
\begin{eqnarray}
&&m_{{H_d^{-}}H_d^{-,*}}=\frac{1}{8}((g_2^2+g_X^2)v_d^2+(-g_X^2+g_2^2)v_u^2+(g_1^2+g_{YX}^2)(-v_u^2+v_d^2)-2g_X^2v_{\bar{\eta}}^2
\nonumber\\&&\hspace{1.8cm}+2(g_{YX}g_X(-v_{\bar{\eta}}^2-v_u^2+v_d^2+v_\eta^2)+g_X^2v_\eta^2)
\nonumber\\&&\hspace{1.8cm}+\frac{1}{2}(2\mid\mu\mid^2+2\sqrt{2}\chi S\Re(\mu\lambda_H^*)+\chi S^2\mid\lambda_H\mid^2,
\end{eqnarray}
\begin{eqnarray}
&&m_{H_d^{-}H_u^{+}}=\frac{1}{2}(2(\lambda_Hl_W^*+B_\mu)+\lambda_H(2\sqrt{2}\chi SM_S^*-v_dv_u\lambda_H^*+v_\eta v_{\bar{\eta}}\lambda_C^*+\sqrt{2}\chi ST_{\lambda_H}))
\nonumber\\&&\hspace{1.8cm}+\frac{1}{4}g_2^2v_dv_u,
\end{eqnarray}
\begin{eqnarray}
&&m_{H_u^{+,*}H_u^{+}}=\frac{1}{8}((-g_X^2+g_2^2)v_d^2+(g_2^2+g_X^2)v_u^2+(g_1^2+g_{YX}^2)(-v_d^2+v_u^2)-2g_X^2v_\eta^2
\nonumber\\&&\hspace{1.8cm}+2(g_{YX}g_X(-v_d^2-v_\eta^2+v_u^2+v_{\bar{\eta}}^2)+g_X^2v_{\bar{\eta}}^2))
\nonumber\\&&\hspace{1.8cm}+\frac{1}{2}(2\mid\mu\mid^2+2\sqrt{2}\chi S\Re(\mu\lambda_H^*)+\chi S^2\mid\lambda_H\mid^2).
\end{eqnarray}
The mass matrix for neutralino in the basis($\lambda_{\tilde{B}}$,$\tilde{W}^0$,$\tilde{H}_d^0$,$\tilde{H}_u^0$,$\lambda_{\tilde{X}}$,$\tilde{\eta}$,$\tilde{\bar{\eta}}$,$\tilde{s}$) is
\begin{eqnarray}
m_{\tilde{\chi}^0}=
\left({\begin{array}{*{20}{c}}
M_1 & 0 & -\frac{g_1}{2}v_d & \frac{g_1}{2}v_u & M_{{BB}^{\prime}} & 0 & 0 & 0 \\
0 & M_2 & \frac{g_2}{2}v_d & -\frac{g_2}{2}v_u & 0 & 0 & 0 & 0 \\
-\frac{g_1}{2}v_d & \frac{g_2}{2}v_d & 0 & m_{{\tilde{H}_u^0}{\tilde{H}_d^0}} & m_{\lambda_{\bar{X}}\tilde{H}_d^0} & 0 & 0 & -\frac{\lambda_Hv_u}{\sqrt{2}} \\
\frac{g_1}{2}v_u & -\frac{g_2}{2}v_u & m_{{\tilde{H}_d^0}{\tilde{H}_u^0}} & 0 & m_{\lambda_{\bar{X}}{\tilde{H}_u^0}} & 0 & 0 & -\frac{\lambda_Hv_d}{\sqrt{2}} \\
M_{{BB}^\prime} & 0 & m_{\tilde{H}_d^0\lambda_{\bar{X}}} & m_{\tilde{H}_u^0\lambda_{\bar{X}}} & M_{BL} & -g_X{v_\eta} & g_Xv_{\bar{\eta}} & 0 \\
0 & 0 & 0 & 0 & -g_X{v_\eta} & 0 & \frac{1}{\sqrt{2}}\lambda_Cv_S & \frac{1}{\sqrt{2}}\lambda_Cv_{\bar{\eta}} \\
0 & 0 & 0 & 0 & g_Xv_{\bar{\eta}} & \frac{1}{\sqrt{2}}\lambda_Cv_S & 0 & \frac{1}{\sqrt{2}}\lambda_Cv_\eta \\
0 & 0 & -\frac{1}{\sqrt{2}}\lambda_Hv_u & -\frac{1}{\sqrt{2}}\lambda_Hv_d & 0 & \frac{1}{\sqrt{2}}\lambda_Cv_{\bar{\eta}} &  \frac{1}{\sqrt{2}}\lambda_Cv_\eta & m_{\tilde{s}\tilde{s}} \\
\end{array}}
\right),
\end{eqnarray}
\begin{eqnarray}
&&m_{{\tilde{H}_d^0}{\tilde{H}_u^0}}=-\frac{1}{\sqrt{2}}\lambda_Hv_S - \mu, ~~ m_{{\tilde{H}_d^0}\lambda_{\bar{X}}}=-\frac{1}{2}(g_{YX}+g_X){v_d},
\nonumber\\&&\ m_{\tilde{H}_u^0\lambda_{\bar{X}}}=\frac{1}{2}(g_{YX}+g_X)v_u, ~~ m_{\tilde{s}\tilde{s}}=2M_s+\sqrt{2}\kappa v_S.
\end{eqnarray}
This matrix is diagonalized by $Z^N$,
\begin{eqnarray}
Z^{N*}m_{{\tilde{\chi}}^0}Z^{N\dag}=m_{{\tilde{\chi}}^0}^{dia}.
\end{eqnarray}
Here are the coupling coefficients for these corresponding vertices:
\begin{eqnarray}
&&H_{W\chi^\pm \chi^0}^L=-\frac{1}{2}g_2(2U_{j1}^{*}N_{i2}+\sqrt{2}U_{j2}^{*}N_{i3}),
\nonumber\\&&
H_{W\chi^\pm \chi^0}^R=-\frac{1}{2}g_2(2N_{i2}^{*}V_{j1}-\sqrt{2}N_{i4}^{*}V_{j2}),
\nonumber\\&&
H^L_{Zll}=\frac{1}{2}(-g_1\cos\theta_W^{\prime}\sin\theta_W+g_2\cos\theta_W\cos\theta_W^{\prime}+g_{YX}\sin\theta_W^{\prime}),
\nonumber\\&&
H^R_{Zll}=-\frac{1}{2}\Big(2g_1\cos\theta_W^{\prime}\sin\theta_W-(2g_{YX}+g_X)\sin\theta_W^{\prime}\Big),
\nonumber\\&&
H_{Z\chi^\pm\chi^\pm}^L=\frac{1}{2}\Big(2g_2U_{i1}^{*}\cos\theta_W\cos\theta_W^{\prime}U_{i1}+U_{i2}^{*}(-g_1\cos\theta_W^{\prime}\sin\theta_W
+g_2\cos\theta_W\cos\theta_W^{\prime}
\nonumber\\&&\hspace{+1.8cm}+(g_{YX}+g_X)\sin\theta_W^{\prime})U_{i2}\Big),
\nonumber\\&&
H_{Z\chi^\pm\chi^\pm}^R=\frac{1}{2}\Big(2g_2V_{i1}^{*}\cos\theta_W\cos\theta_W^{\prime}V_{i1}+V_{i2}^{*}(-g_1\cos\theta_W^{\prime}\sin\theta_W
+g_2\cos\theta_W\cos\theta_W^{\prime}
\nonumber\\&&\hspace{+1.8cm}+(g_{YX}+g_X)\sin\theta_W^{\prime})V_{i2}\Big),
\nonumber\\&&
H_{H\chi^\pm \chi^0}^L=\frac{1}{2}\Big(-2g_2V_{i1}^{*}N_{j4}^{*}Z_{k2}^{+}-V_{i2}^{*}(2\lambda_HN_{j8}^{*}Z_{k1}^{+}+\sqrt{2}(g_1N_{j1}^{*}+g_2N_{j2}^{*}
\nonumber\\&&\hspace{+1.8cm}+(g_{YX}+g_X)N_{j5}^{*})Z_{k2}^{+})\Big),
\nonumber\\&&
H_{H\chi^\pm \chi^0}^R=\frac{1}{2}\Big(-2g_2U_{i1}^{*}N_{j3}Z_{k1}^{+}+U_{i2}(-2\lambda_H^{*}N_{j8}Z_{k2}^{+}+\sqrt{2}(g_1N_{j1}+g_2N_{j2}
\nonumber\\&&\hspace{+1.8cm}+(g_{YX}+g_X)N_{j5})Z_{k1}^{+})\Big),
\nonumber\\&&
H_{h_0\chi^\pm\chi^\pm}^L=-\frac{1}{\sqrt{2}}\Big(g_2U_{j1}^{*}V_{j2}^{*}Z_{k2}^{H}+U_{j2}^{*}(g_2V_{j1}^{*}Z_{k1}^{H}+\lambda_HV_{j2}^{*}Z_{k5}^{H})\Big),
\nonumber\\&&
H_{h_0\chi^\pm\chi^\pm}^R=-\frac{1}{\sqrt{2}}\Big(g_2U_{j1}V_{j2}Z_{k2}^{H}+U_{j2}(g_2V_{j1}Z_{k1}^{H}+\lambda_H^{*}V_{j2}Z_{k5}^{H})\Big)
\end{eqnarray}

\begin{eqnarray}
&&H_{Z\chi^0\chi^0}^L=-\frac{1}{2}\Big(N_{i3}^{*}(g_1\cos\theta_W^{\prime}\sin\theta_W+g_2\cos\theta_W\cos\theta_W^{\prime}-(g_{YX}+g_X)\sin\theta_W^{\prime})N_{i3}
\nonumber\\&&\hspace{+1.8cm}-N_{i4}^{*}(g_1\cos\theta_W^{\prime}\sin\theta_W+g_2\cos\theta_W\cos\theta_W^{\prime}-(g_{YX}+g_X)\sin\theta_W^{\prime})
\nonumber\\&&\hspace{+1.8cm}+2(-g_X\sin\theta_W^{\prime})(N_{i6}^{*}N_{i6}-N_{i7}^{*}N_{i7})\Big),
\nonumber\\&&
H_{Z\chi^0\chi^0}^R=\frac{1}{2}\Big(N_{i3}^{*}(g_1\cos\theta_W^{\prime}\sin\theta_W+g_2\cos\theta_W\cos\theta_W^{\prime}-(g_{YX}+g_X)\sin\theta_W^{\prime})N_{i3}
\nonumber\\&&\hspace{+1.8cm}-N_{i4}^{*}(g_1\cos\theta_W^{\prime}\sin\theta_W+g_2\cos\theta_W\cos\theta_W^{\prime}-(g_{YX}+g_X)\sin\theta_W^{\prime})
\nonumber\\&&\hspace{+1.8cm}+2(-g_X\sin\theta_W^{\prime})(N_{i6}^{*}N_{i6}-N_{i7}^{*}N_{i7})\Big),
\nonumber\\&&
H_{h_0\chi^0\chi^0}^L=\frac{1}{2}\Big(Z_{k1}^{H}[g_1(N_{i1}N_{i3}+N_{i3}N_{i1})-g_2(N_{i2}N_{i3}+N_{i3}N_{i2})
\nonumber\\&&\hspace{+1.8cm}+(g_{YX}+g_X)(N_{i5}N_{i3}+N_{i3}N_{i5})+\sqrt{2}\lambda_H(N_{i4}N_{i8}+N_{i8}N_{i4})]^{*}
\nonumber\\&&\hspace{+1.8cm}+Z_{k2}^{H}[-g_1(N_{i1}N_{i4}+N_{i4}N_{i1})g_2(N_{i2}N_{i4}+N_{i4}N_{i2})
\nonumber\\&&\hspace{+1.8cm}-(g_{YX}+g_X)(N_{i5}N_{i4}+N_{i4}N_{i5})+\sqrt{2}\lambda_H(N_{i3}N_{i8}+N_{i8}N_{i3})]^{*}
\nonumber\\&&\hspace{+1.8cm}+Z_{k3}^{H}[2g_X(N_{i5}N_{i6}+N_{i6}N_{i5})-\sqrt{2}\lambda_C(N_{i7}N_{i8}+N_{i8}N_{i7})]^{*}
\nonumber\\&&\hspace{+1.8cm}+Z_{k4}^{H}[-2g_X(N_{i5}N_{i7}+N_{i7}N_{i5})-\sqrt{2}\lambda_C(N_{i6}N_{i8}+N_{i8}N_{i6})]^{*}
\nonumber\\&&\hspace{+1.8cm}+Z_{k5}^{H}[\sqrt{2}\lambda_H(N_{i3}N_{i4}+N_{i4}N_{i3})-\sqrt{2}\lambda_C(N_{i7}N_{i6}+N_{i6}N_{i7})
\nonumber\\&&\hspace{+1.8cm}-2\sqrt{2}\kappa N_{i8}N_{i8}]^{*}\Big),
\nonumber\\&&
H_{h_0\chi^0\chi^0}^R=\frac{1}{2}\Big(Z_{k1}^{H}[g_1(N_{i1}N_{i3}+N_{i3}N_{i1})-g_2(N_{i2}N_{i3}+N_{i3}N_{i2})
\nonumber\\&&\hspace{+1.8cm}+(g_{YX}+g_X)(N_{i5}N_{i3}+N_{i3}N_{i5})+\sqrt{2}\lambda_H(N_{i4}N_{i8}+N_{i8}N_{i4})]
\nonumber\\&&\hspace{+1.8cm}+Z_{k2}^{H}[-g_1(N_{i1}N_{i4}+N_{i4}N_{i1})g_2(N_{i2}N_{i4}+N_{i4}N_{i2})
\nonumber\\&&\hspace{+1.8cm}-(g_{YX}+g_X)(N_{i5}N_{i4}+N_{i4}N_{i5})+\sqrt{2}\lambda_H(N_{i3}N_{i8}+N_{i8}N_{i3})]
\nonumber\\&&\hspace{+1.8cm}+Z_{k3}^{H}[2g_X(N_{i5}N_{i6}+N_{i6}N_{i5})-\sqrt{2}\lambda_C(N_{i7}N_{i8}+N_{i8}N_{i7})]
\nonumber\\&&\hspace{+1.8cm}+Z_{k4}^{H}[-2g_X(N_{i5}N_{i7}+N_{i7}N_{i5})-\sqrt{2}\lambda_C(N_{i6}N_{i8}+N_{i8}N_{i6})]^{*}
\nonumber\\&&\hspace{+1.8cm}+Z_{k5}^{H}[\sqrt{2}\lambda_H(N_{i3}N_{i4}+N_{i4}N_{i3})-\sqrt{2}\lambda_C(N_{i7}N_{i6}+N_{i6}N_{i7})
\nonumber\\&&\hspace{+1.8cm}-2\sqrt{2}\kappa N_{i8}N_{i8}]\Big).
\end{eqnarray}

\end{document}